# Analyzing the Relationship between Project Productivity and Environment Factors in the Use Case Points Method

Mohammad Azzeh
Department of Software Engineering
Applied Science University
Amman, Jordan POBOX 166
m.y.azzeh@asu.edu.jo

Ali Bou Nassif
Department of Electrical and Computer Engineering
University of Sharjah
Sharjah, UAE
anassif@sharjah.ac.ae

**Abstract.**
Project productivity is a key factor for producing effort estimates from Use Case Points (UCP), especially when the historical dataset is absent. The first versions of UCP effort estimation models used a fixed number or very limited numbers of productivity ratios for all new projects. These approaches have not been well examined over a large number of projects so the validity of these studies was a matter for criticism. The newly available large software datasets allow us to perform further research on the usefulness of productivity for effort estimation of software development. Specifically, we studied the relationship between project productivity and UCP environmental factors, as they have a significant impact on the amount of productivity needed for a software project. Therefore, we designed four studies, using various classification and regression methods, to examine the usefulness of that relationship and its impact on UCP effort estimation. The results we obtained are encouraging and show potential improvement in effort estimation. Furthermore, the efficiency of that relationship is better over a dataset that comes from industry because of the quality of data collection. Our comment on the findings is that it is better to exclude environmental factors from calculating UCP and make them available only for computing productivity. The study also encourages project managers to understand how to better assess the environmental factors as they do have a significant impact on productivity.

## 1. Introduction

Software effort estimation is necessarily required at the inception phase in order to bid for a software project and assign human resources [1] [2] [3] [4] [45]. Since very little data are known at this stage, managers turn their attention to software size metrics, which can help in providing a general sense about the most likely effort of a software project. Functional size metrics, such as Function Points (FP) [5] and Use Case Points (UCP) [6], are well suited measures for such a problem. The FP has long been studied and examined as a more efficient size measure [7] [8]. Nevertheless, measuring FP is tedious and requires intensive care, especially counting FP components such as internal, external, and query functions, in addition to evaluating environmental and complexity factors [3] [9].

More recently, the UCP method has attracted more attention from researchers as a competitive sizing method to FP at the early phases of software development. The UCP was first introduced by Karner [6] in 1993 to estimate the size of object-oriented software projects. The basic idea of constructing UCP was inspired by the FP method. The typical input of UCP is the UML use case diagram, which describes the functional requirements in UML annotations such as actors, use cases, and various kinds of relations [10]. The UCP is computed by converting the elements of use case diagrams into their corresponding size metrics through a well-defined procedure. The accuracy of the UCP size metric mainly depends on the level of detail that is attached to the use case diagram [4] [11]. So it is risky to use this method without providing adequate information. However, in spite of its simplicity in comparison with its counterpart FP, the UCP is not impressive as an efficient effort estimation within software organizations, because there is no consensus on how to translate the UCP size into a corresponding amount of effort. In the early version of UCP, Karner [6] recommended multiplying the calculated UCP with a pre-determined productivity ratio to compute the most likely effort, as shown in Equation 1. This approach is rather straightforward but risky if the manager does not have enough knowledge about the productivity of his/her project. Software productivity is defined as a ratio between effort and size [12] [13] [14]. This definition, rather simple is a matter of some debate because there is no standard on how to represent that relationship [12]. Since the productivity used in this paper is computed from completed projects, we call it Project Delivery Rate (PDR). The terms project productivity and PDR will be used interchangeably throughout this paper. Some authors interpret the PDR as $size/effort$, so when size increases the productivity also increases [10] [12]. Other studies interpret PDR as $effort/size$, so when the size decreases the required productivity increases [1] [13]. The effort estimation model is usually constructed based on the productivity interpretation. For example, Equation 1 is used to estimate the effort when PDR is defined as effort/size. In contrast, the nonlinear effort estimation model is used when PDR is defined as size/effort, as shown in Equation 10. The term 'Productivity' in Equation 10 represents the productivity of the team who is developing the project. Nevertheless, computing the software productivity must be made before predicting the effort; this depends on many factors such as amount of code reuse, type of process model, team communications, and number of project deliverables [15]. Adopting good development practices may increase the productivity but does not always do so because of circumstances outside the control of the software development team [13].

$$Effort = Productivity \times UCP \tag{1}$$

Owing to its notable predictive accuracy, project productivity has long been involved as a cost driver in computing efforts from UCP when there are no available historical datasets. Although several studies have been conducted to compute productivity before predicting the software effort [6] [10] [16], most of them use very limited levels of productivity ratio. These approaches do not take the project characteristics into consideration. Karner [6] suggests the use of only one level of productivity, which is 20 person-hours/UCP. This value was obtained after analyzing the post productivity for three complete industrial projects. Other studies attempted to predict the productivity from environmental factors, as they found that these factors exhibit strong indications about project productivity. In this direction, Schneider and Winters (S&W) [16] proposed three levels of productivity ratio (fair=20, low=28, very low=36) person-hours/UCP based on performing a specific analysis on environmental factors. Similarly, Nassif et al. [10] proposed four levels of productivity ratio based on analyzing environmental factors and using an expert-based fuzzy model. The productivity obtained is then involved with their nonlinear regression model to compute the effort. The key difference between Nassif's model and the first two models is that the Karner model and the S&W model presume that the relationship between effort and size is linear. This is incorrect from the viewpoint of Nassif et al. [10], because when software size increases, the number of team members required to develop this software increases. Furthermore, when the team becomes larger, communication overhead will increase and this requires additional effort [12].

Having discussed the basic effort estimation models that use productivity and UCP as predictors, we can figure out the following findings regarding the project productivity. Karner's assumption is not a realistic approach because it does not take into consideration the differences between software projects in terms of their types, environments, and complexity. The S&W model attempts to classify a software project into one of three categories based on studying and counting environmental factors using a specific counting procedure. The project classification procedure only depends on counting the factors, ignoring the influence of their values. Similarly, the productivity classification procedure proposed by Nassif et al. [10] was based on the prod-sum variable, using an expert based model. Perhaps if the authors used some clustering techniques based on the collected data, the model would work better. So we believe that the productivity computation should be flexible and adjustable when data is available. The flexibility means that the productivity must be affected by the UCP factor assessment. The adjustability means the productivity of one project should be adjusted based on the productivities of the historical projects. Since very little has been done to study the benefit that can be obtained from analyzing environmental factors to predict productivity, the new available large software repository empirically allows the ability to perform further research on productivity and UCP for early effort estimation. This paper attempts to examine the following research questions:

RQ1: Is there a useful relationship between UCP environmental factors and software project productivity? If that relationship already exists, what is the best way to model that relationship?

RQ2: Which is better for UCP-based effort estimation, the dynamic or the static productivity ratio?

To answer these questions, we designed four studies to investigate how we can efficiently learn and predict project productivity from environmental factors. More details about environmental factors are discussed in Section 2. The studies conducted by [10] and [16] showed that the environmental factors can work as good indicators for software productivity since they reflect the team workload within the software project. Continuing along that direction, the first study was designed to investigate how we can benefit from the S&W procedure to further improve the productivity prediction from environmental factors. For that purpose, we used classification algorithm with class decomposition to build a learning model which can predict project productivity from environment factors. Similarly, the second study attempted to decompose the productivity variable into several classes and give each class a specific label. Then, the training data were entered into a classification algorithm to learn productivity for the new test project. The difference between the first and second study is that in the first study the training projects were initially labeled based on the S&W procedure while in the second study we used clustering to create labels. The third study was designed to build a multiple regression model between productivity and environment factors. The fourth study was to apply analogy-based estimation with regression to mean on the environment factors and productivity. In these studies, the predicted productivity was multiplied with UCP, as shown in Equation 1, in order to estimate the most likely effort.

The remainder of this paper is organized as follows: Section 2 introduces the UCP method. Section 3 presents related work. Section 4 presents experiment setup and methodology. Section 5 describes the evaluation measures. Section 6 introduces the dataset description and productivity analysis. Section 7 shows the empirical results and discussion. Section 8 presents threats to validity and, finally, Section 9 ends the paper with the conclusions of the study.

## 2. An overview of Use Case Points

The UCP estimation method was proposed by Karner [6] to predict the size of object-oriented software projects. The UCP is computed by translating the components of the use case diagram into corresponding size metrics, which can be later adjusted by complexity and technical factors. The use case diagram is the typical input of UCP, which describes functional requirements. This makes the UCP a good choice at the early stage of software development. The predictive accuracy of computing UCP primarily depends on the amount of description that is attached with use cases. So it is mandatory to avoid free style use case description and follow a specific guideline [17] [18]. The recommended semantic of use case description commonly includes the following sections: title, main actor(s), pre- and post-conditions, main scenario, alternative scenario, and exceptions. However, there is a lack of clear definition of how to interpret the description of use case into corresponding steps, paths, or transactions.

The UCP is calculated by computing four components. The first component is obtained by computing the Unadjusted Actor Weights (UAW) which can be done by performing Equation 2. The actors are usually classified into three categories (simple, average, and complex) according to their difficulties. The simple actors category represents systems with a defined application programming interface (API). Average actors are actors that can communicate through a protocol such as TCP/IP, FTP, and HTTP or actors that are data stores (Files, RDBMS). Complex actors represent persons interacting through GUI or web pages [1].

$$UAW = 1 \times a_s + 2 \times a_a + 3 \times a_c \tag{2}$$

Where $a_s, a_a, a_c$ are the numbers of simple, average, and complex actors, respectively.

The second component is the Unadjusted Use Cases (UUCW), which can be calculated as shown in Equation 3. Firstly, the use cases are classified into three classes (simple, average, and complex) based on the number of transactions in the use case description. A transaction is defined as a stimulus and response occurrence between the actor and the system [2]. A simple use case contains less than or equal to three transactions, an average use case contains between four and seven transactions and finally, the complex use case contains more than seven transactions.

$$UUCW = 5 \times uc_s + 10 \times uc_a + 15 \times uc_c \tag{3}$$

Where $uc_s, uc_a, uc_c$ are the numbers of simple, average, and complex use cases, respectively.

The summation of UAW and UUCW is called Unadjusted Use Case Points (UUCP) as shown in Equation 4.

$$UUCP = UAW + UUCW \tag{4}$$

To compute UCP, the UUCP must be adjusted by Technical Complexity and Environmental factors. Technical Complexity Factors (TCF) consist of thirteen factors that have great influence on project performance. Each factor can take a value between 0 and 5 inclusive, which represents the influence of that factor, where the value zero means no influence and the value five represents strong influence. Each factor is also multiplied by a predetermined weight from Table 1. The technical complexity factor is then computed as shown in Equation 5.

TABLE 1. Technical complexity adjustment factors

|  | Technical Factor | Weight($fw$) |
|---|---|---|
| $f_1$ | Distributed System | 2 |
| $f_2$ | Response Objective | 2 |
| $f_3$ | End User Efficiency | 1 |
| $f_4$ | Complex Processing | 1 |
| $f_5$ | Reusable Code | 1 |
| $f_6$ | Easy to Install | 0.5 |
| $f_7$ | Easy to Use | 0.5 |
| $f_8$ | Portable | 2 |
| $f_9$ | Easy to Change | 1 |
| $f_{10}$ | Concurrent | 1 |
| $f_{11}$ | Security Features | 1 |
| $f_{12}$ | Access for Third Parties | 1 |
| $f_{13}$ | Special Training Required | 1 |

$$TCF = 0.6 + \left(0.01 \times \sum_{i=1}^{13}(f_i \times fw_i)\right) \qquad (5)$$

where $f_i$ is the value of influence of factor $i$, and $fw_i$ is the weight associated with factor $i$ from Table 1.

Environmental factors consist of eight factors that have great influence on project productivity, as shown in Table 2. Since the paper is about environmental factors, we describe these factors as follows [16]:

1. *Familiar with RUP*: This factor reflects the familiarity of the team with the standard of Rational Unified Process (RUP) framework. Higher experience leads to better productivity and less effort.
2. *Application Experience*: This factor reflects the experience of the team with the project type. Higher numbers represent more experience, thus less effort.
3. *Object-Oriented experience*: This factor reflects the experience of the team with Object-Oriented Programming and development. Higher numbers represent more experience and less effort.
4. *Lead Analyst Capability*: This factor represents the capability and knowledge of the person who is responsible for the requirements. Higher numbers represent increased skill and knowledge, thus less effort.
5. *Motivation*: This factor represents the degree of developer motivation. The motivated team leads to better productivity.
6. *Stable Requirements*: This factor measures the stability of requirements changes. Large stability leads to less effort.
7. *Part-Time Staff*: This factor indicates the level of outside developers and consultants needed to accomplish a software project. This has a negative effect on the productivity and effort.
8. *Difficult Programming Language*: This factor represents the difficulty of the programming language used. This factor also has a reverse effect because it will increase effort.

Each factor can take a value between 0 (no influence) and 5 (strong influence). Also, each factor has a predefined weight. The last two factors have negative weights to indicate that these factors have reverse effects. The Environmental Factor (EF) is computed as shown in Equation 6.

$$EF = 1.4 - \left(0.03 \times \sum_{i=1}^{8}(env_i \times ew_i)\right) \qquad (6)$$

where $env_i$ is the value of influence of factor $i$, and $ew_i$ is the weight associated with factor $i$ from Table 2.

Finally, the adjusted use case points UCP is calculated by multiplying the UUCP with TCF and EF as shown in Equation 7.

$$UCP = UUCP \times TCF \times EF \qquad (7)$$

TABLE 2. Environmental factors

| | Environment Factor | Weight($ew$) |
|---|---|---|
| $env_1$ | Familiar with RUP | 1.5 |
| $env_2$ | Application experience | 0.5 |
| $env_3$ | Object-Oriented experience | 1 |
| $env_4$ | Lead Analyst capabilities | 0.5 |
| $env_5$ | Motivation | 1 |
| $env_6$ | Stable requirements | 2 |
| $env_7$ | Part-time staff | -1 |
| $env_8$ | Difficult programming language | -1 |

## 3. Related Work

Estimating the most likely software effort from UCP was the core subject of many studies conducted to improve early effort estimation [6] [19] [20] [21] [22] [23] [24] [25] [44]. Since the amount of collected data at the inception phase of the software development life cycle were not sufficient, most UCP estimation models tended to involve productivity as the second predictor to estimate the effort, as shown in Equation 1. In this context, the first study conducted by Karner [6] recommended the use of a fixed productivity ratio, which is 20 person-hours/UCP, as shown in Equation 8. This value has been found after studying and analyzing three complete industrial software projects. The productivity obtained in this study is called post productivity since it was measured from complete projects. Karner generalized his findings and recommended using that productivity ratio as the default value for software organizations that do not have historical datasets. It is quite astonishing that this assumption became the basis for the majority of UCP-based estimation models irrespective of its validity or suitability for some organizations. Thus, it might work efficiently with a software organization that does not have historical data. But when an organization has sufficient historical data, it is recommended to calibrate the productivity by using that historical data. We believe that using a fixed productivity ratio for all software projects is not a typical solution because each software project has different characteristics and has been developed in different environments.

$$Effort = 20 \times UCP \tag{8}$$

Schneider and Winters [16] have performed more analyses on UCP metrics, especially environmental factors. They found that the environmental factors could work as indicators of the productivity ratio. Based on that, they proposed three levels of productivity value: fair (20 person-hours/UCP), low (28 person-hours/UCP), and very low (36 person-hours/UCP). The classification procedure was achieved by analyzing the environmental factors. Basically, one should count the number of factors that have influence values of less than 3 from the set ($env_1$ to $env_6$) and count the number of factors that have influence values larger than 3 from the set ($env_7$ to $env_8$). The efficiency is evaluated based on total count. If the total count is less than or equal to 2 then the efficiency is fair. If the total count is between 3 and 4 (inclusive) the efficiency is low, and finally if the total count is greater than 4 the efficiency is very low. For a fair efficiency project, the productivity is 20 person-hours per UCP. For a low efficiency project, the productivity is 28 person-hours per UCP. For a very low efficiency project, the productivity is 36 person-hours per UCP. The general form for this model is shown in Equation 9. It is important to note that the environmental factors are not included in computing UCP.

$$Effort = \begin{cases} 20 \times UCP & total\_Count \leq 2 \\ 28 \times UCP & 3 \leq total\_Count \leq 4 \\ 36 \times UCP & total\_Count > 4 \end{cases} \tag{9}$$

This approach has limited levels of productivity with no flexibility to adjust them based on the project characteristics. On the other hand, Nassif et al. [10] proposed a non-linear relationship between effort and UCP using productivity as the input coefficient, as shown in Equation 10. Nassif and his colleagues assumed that the relationship between effort and UCP follows an exponential curve. The factors α and β are determined after analyzing the training dataset. In the original model, the values were α=8.16 and β=1.17. The productivity was computed based on finding $prod\_sum = \sum_{i=1}^{8}(env_i \times w_i)$ from the environment factors. Then this value was entered into a fuzzy model that converts the $prod\_sum$ into productivity value using some predefined rules. It is important to note that the term productivity represents the team productivity and it is interpreted as size/effort in this model.

$$Effort = \frac{\alpha}{Productivity} \times UCP^{\beta} \tag{10}$$

Anda et al. [17] [18] [23] found that the UCP effort estimate depends heavily on the technical and environmental factors. They recommended that the UCP model can be tailored by adjusting the environmental factors based on the organization type. However, other studies used machine learning algorithms and fuzzy systems to build effort prediction models based only on UCP. In this context, Nassif et al. [26] [27] [28] used Mamdani and Sugeno fuzzy models to treat the abrupt change in the complexity weights of the use cases. They also proposed another model based on the Treeboost algorithm [22]. Braz et al. [19] introduced the metric Fuzzy Use Case Size Points (FUSP) that enabled gradual classifications in the estimation by using fuzzy numbers. Neural network-based models [20] [21] [29] have also been used to predict the software effort. The inputs of these neural network models include software size in UCP and other quality attributes such as complexity. On the other hand, some studies focused on the construction of UCP; for example, Robiolo [2] proposed two metrics called transactions and paths that capture software size and complexity. The two metrics reduce the estimation error. Ochodek et al. [1] [30] proposed a new method for measuring use-case complexity that is called TTPoints. It includes knowledge regarding semantics of transactions, numbers of business objects, and interacting actors. They also proposed simplification to the UCP model by eliminating the weight of the actors because it is insignificant with respect to the use case weights. Recently, Silhavy et al. [46] conducted various experiments to study the significant impact of UCP variables. They found that all UCP variables are significant in computing size and effort, but with different degrees.

Having reviewed the closely related literature, we assert that the usefulness relationship between productivity and environment factors has not been sufficiently explored. Thus, this work augments the existing body of knowledge in the area of UCP effort estimation with some proposed ideas for productivity prediction.

## 4   Experimental Setup

The goal of the experiments we conducted was to examine the usefulness of environmental factors for predicting productivity. To achieve that goal, we designed four studies based on large datasets collected from a software organization and university students. It is necessary to bear in mind that the UCP for each project was computed without involving environmental factors (i.e., only UAW, UUCW, and TCF were included). The UCP construction is out of the scope of this paper; we are only interested in predicting project productivity from environmental factors and then estimating the effort from UCP and predicted productivity. Other prediction models based on UCP, such as Fuzzy models and regression models that do not use productivity ratio as input for effort estimation, are out of the scope of this paper. The other forms of the UCP size metric are also out of scope.

The philosophy of the experiments used in this paper is illustrated in Figure 1. Across all studies, the environmental factors will be used as input for all constructed productivity prediction models. The resulting productivity and estimated UCP will be used as input for an effort prediction model, as explained in Equation 1. However, the productivity prediction model for each study is thoroughly explained in the upcoming subsections.

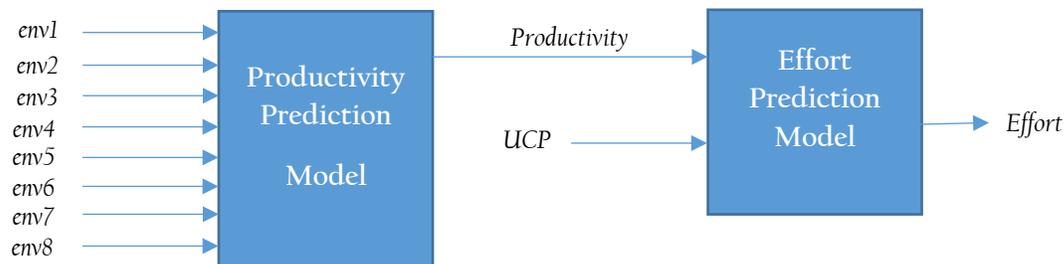

Figure 1. Illustration of the methodology used in the four studies.

The leave-one-out cross validation has been used to validate each model. Although some authors recommend n-Fold cross validation, the leave-one-out cross validation has been used in a deterministic procedure that can be exactly repeated by any other research with access to a particular dataset [35]. It also ensures that any prediction model is constructed from the same set of training projects. Based on literature, the leave-one-out cross validation produces lower bias estimates and higher variance than n-Fold cross validation, because the method needs to learn from a large number of examples and more tests are conducted [3]. The mechanism of leave-one-out cross validation is executed as follows: in each run, one project is held out as a test set and the remaining projects act as a training set. The prediction model is developed on the training set while the test set is used to evaluate the model. The error measures are calculated for each test instance. This procedure is continued until all projects within the dataset act as test projects.

### 4.1 The first study

The first study was designed to further improve the S&W [16] model by adding learning capability. The main limitation of the S&W model is that it only uses three labels to classify projects and this is not sufficient from the managerial point of view. Also the classification procedure is not well defined because it only depends on a counting mechanism with a certain threshold. Nevertheless, the predictive accuracy of S&W proved to be significantly better than the original Karner model [6]. Therefore, we decided to use a classification algorithm and clustering technique to produce additional class labels that were more coherent than only three class labels. Figure 2 illustrates the validation and construction procedure of the proposed model in the first study (hereafter M1). We first applied the S&W model on the training environmental factors to classify projects into three classes (i.e., fair, low, and very low). Each class (including both environmental factors and actual productivity) was then decomposed using a clustering algorithm to create more labels within an initial class (i.e., clusters). This step is very important in order to provide better representation of the data. The center of each new cluster will be the most represented value for that new class label. The training environment factors and new labeled productivity that were obtained were then entered into a classification algorithm to learn the productivity label for a new test project. If test project $x$ belongs to class $c$, then the productivity center of that class is used as predicted productivity for that test project. The estimated productivity is then multiplied with the UCP of the test project to compute the estimated effort using Equation 1.

In this study, we used Random Forests (RF) classification algorithm [31] because it has been proven to be one of the most accurate algorithms [32]. RF is an ensemble learning technique that has been successfully used for classification and regression. The algorithm works as a set of independent classifiers (typically decision trees), where each classifier casts a vote for a particular instance in a dataset, and then majority voting is considered to determine the final class label [31]. The setup parameters of the RF algorithm that is used in this study are described in Table 3.

TABLE 3. Parameter setups of RF

| Parameter | value |
|---|---|
| Number of trees | 10 |
| Min Leaf | 3 |
| Prune | true |
| minimum node size to split | 1 |

For class-decomposition we used the k-means clustering algorithm [33] aiming at minimizing the within-cluster sum of squares for each group of instances that belong to the same class label [33]. The main challenge with using k-means is how to determine the optimum number of clusters for particular data. To do so, we applied k-means repetitively on the whole dataset with a varying number of $k$ clusters from 2 to ($m/3$), where $m$ is the number of projects in the initial class label, and choose the best $k$ that produces minimum validity value. Equation 11 shows the clustering validity measure used in this paper, which is based on minimizing the distances between observations within the same cluster and maximizing the distance between different clusters. The optimum $k$ value will be used later to cluster class labels across all training sets.

$$validity = \frac{\frac{1}{m}\sum_{i=1}^{k}\sum_{x \in c_i}||x - z_i||}{\min_{i \neq j}(||z_i - z_j||^2)} \quad (11)$$

Where $z_i$ and $z_j$ are the centers of $i^{th}$ and $j^{th}$ clusters, $k$ is the number of clusters, $m$ is the number of observations. $x$ is an observation that belongs to the $i^{th}$ clusters ($c_i$).

```
1:   [Label₁, Label₂, Label₃]=S&W(data)
2:   for j=1 to 3
3:       M←Size(Labelⱼ)
4:       minValidity←inf
5:       for k=1 to m/3
6:           Validity←evaluate(k-means(Labelⱼ,k))
7:           If(validity < minValidity)
8:               minValidity←Validity
9:               bestK←k
10:          end
11:      end
12:      for i=1 to n
13:          test← Labelⱼ(i)
14:          Train← Labelⱼ(~i)
15:          [newLabel] ←k-means(train,bestK)
16:          ClassificationModel←RF(newLabel)
17:          Productivity(i) ←predict(ClassificationModel, test)
18:          Effort(i) ←Productivity(i) × UCP(i)
19:      end
20:  end
```

Figure 2. Algorithm of the proposed model of the first study

## 4.2 The Second Study

The purpose of this study was to examine the distribution of the training productivity variable in order to create coherent class labels. The labels and values of environmental factors that were discovered will be entered into the RF classification algorithm in order to learn productivity from environmental factors. Figure 3 illustrates the validation and construction procedure of the proposed model from the second study (hereafter M2). Firstly, the training productivity variable was only clustered to $k$ labels using the k-means algorithm. For each label, the center productivity obtained will be the representative value for that label. The optimum number of clusters was found by repetitively applying k-means to the whole dataset with a changing value of k every time, and we chose the value that produces minimum validity, as explained in section 4.1. Then we entered the training dataset (including environmental factors and discovered productivity labels) into RF classification algorithm to learn from the training data. When a new test project is to be estimated we first entered its environmental factors values into the generated classification model to guess to which label the new project belongs. Based on predicting the label, we took the centroid of that label as the predicted productivity that will be multiplied with the UCP of the test project in order to estimate its effort.

```
1:   n<-Size(data)
2:   minValidity←inf
3:   for k=1 to n/3
4:       Validity←evaluate(k-means(data , k))
5:       If(validity < minValidity)
6:           minValidity←Validity
7:           bestK← k
8:       End
9:   End
10:  for i=1 to n
11:      test← data(i)
12:      Train← data(~i)
13:      [newLabel] ←k-means(train,bestK)
14:      ClassificationModel←RF(newLabel)
15:      Productivity(i) ← predict(ClassificationModel, test)
16:      Effort(i) ← Productivity(i) × UCP(i)
17:  End
```

Figure 3. Algorithm of the proposed model of the Second study

## 4.3 The Third Study

In this study, we used the stepwise regression algorithm in order to build a productivity prediction model. The predictors are environmental factors of training projects, while the response variable is the productivity of training projects. The process started with an empty set of factors, then systematically added the most significant variable or removes the least significant variable during each step. The final model will only include the statistically significant factors, but not necessarily the highest $R^2$. The values of each environmental factor were first tested for normality. If the factor was not normal, it was transformed to normal distribution using the Cox-box method [41]. The general model of the stepwise regression, which will be called M3 hereafter, is shown in Equation 12. The predicted productivity is then multiplied by the test UCP in order to estimate the effort as shown in Equation 1.

$$productivtiy = coeff + \sum_{i=1}^{8} \beta_i \times env_i \qquad (12)$$

Where $coeff$ is the intersection coefficient, $\beta_i$ is the coefficient of each environment factor if it is significant, otherwise it is replaced by zero.

## 4.4 The Fourth Study

The Fourth study was designed to examine the usefulness of Analogy-Based Estimation (ABE) [34] [35] with the Regression to Mean (R2M) method [36] to retrieve and calibrate the productivity of the test project based on its neighbors. The ABE is a well-defined method based on the idea of retrieving by similarity. Thus, for each new project, the ABE uses a similarity measure to find the closest analogies within historical data. The ABE model used in this study is the original method that was proposed by Shepperd and Schofield [34], which uses Euclidean distance as a similarity measure and only one nearest analogy is retrieved. The input features for ABE are environmental factors and the output is the productivity. So the ABE attempts to retrieve the closest productivity for the new project based on the similarity degree between its values of environmental factors and those of training projects. The closest productivity is then adjusted by the R2M method, as shown in Equation 13. Jørgensen et al. [36] first applied R2M to ABE in order to estimate the effort from FP. This method assumes that if the nearest projects have extreme productivity values, then the productivity value of the new project should be adjusted to bring it closer to the average productivity values of training projects. However, it is recommended to refer to [37] [38] for more details. The predicted productivity is then multiplied by the test UCP in order to estimate the effort, as shown in Equation 1. The proposed model from this study is called M4 hereafter.

$$Productivtiy = Productivtiy_i + (h - Productivtiy_i) \times (1 - r) \tag{13}$$

where $Productivtiy_i$ is the productivity of $i^{th}$ nearest analogy, $h$ is the average productivity of the training projects, and $r$ is the historical correlation between the non-adjusted analogy based productivity and the actual productivity.

## 5 Evaluation measures

The conventional accuracy measures, such as Magnitude of Relative Error (MRE) and their derived measures Mean Magnitude Relative Error (MMRE) and Performance indicator (*pred*), have been a matter of debate in software engineering literature [39] [40]. The basic measure (i.e., MRE) is regarded as a biased measure because it yields an asymmetry distribution [39]. Thus, in this paper we used alternative measures that produce unbiased and symmetry distribution such as Absolute Error (AE), Mean Absolute Error (MAE), Mean Balanced Relative Error (MBRE), and the Mean Inverted Balanced Relative Error (MIBRE), as shown in Equations 14 to 17 respectively [43].

$$AE_i = |e_i - \hat{e}_i| \tag{14}$$

$$MAE = \frac{\sum_{i=1}^{n} AE_i}{n} \tag{15}$$

$$MBRE = \frac{1}{n} \sum_{i=1}^{n} \frac{AE_i}{min(e_i, \hat{e}_i)} \tag{16}$$

$$MIBRE = \frac{1}{n} \sum_{i=1}^{n} \frac{AE_i}{max(e_i, \hat{e}_i)} \tag{17}$$

Where $e_i$ and $\hat{e}_i$ are the actual and estimated effort of an observation.

On the other hand, we also used two measures that were proposed by Shepperd and MacDonell [40] for evaluating prediction models. These two measures are the Standardized Accuracy (SA) and Effect Size (Δ) as shown in Equations 18 and 19 respectively. The SA measure is mainly used to test whether the prediction model really surpasses a baseline of random guessing and produces meaningful predictions [40]. In other words, it is a ratio of how much better a given prediction model is than random guessing. The Δ is used to examine whether the predictions of a model are generated by chance, and to show the percentage of improvements over random guessing, since the statistical significance test is not so informative if both predictions models are really different[40]. The value of Δ is interpreted in terms of the categories of small (0.2), medium (0.5), and large (0.8) where a value larger than or equal to 0.5 is considered better [40]. In addition to the above evaluation measures, we used Wilcoxon sum rank test for statistical significance test among different prediction models.

$$SA = 1 - \frac{MAE}{\overline{MAE}_{po}} \tag{18}$$

$$\Delta = \frac{MAE - \overline{MAE}_{po}}{SP_o} \tag{19}$$

Where $\overline{MAE}_{po}$ is the mean value of a large number runs of random guessing. This is defined as, predict a $\hat{e}_i$ for the target case $t$ by randomly sampling (with equal probability) over all the remaining $n - 1$ cases and take $e_t = e_r$ where $r$ is drawn randomly from $1...n \wedge r \neq t$. This randomization procedure is robust since it makes no assumptions and requires no knowledge concerning population. $SP_o$ is the sample standard deviation of the random guessing strategy.

## 6 Dataset

The previous models on UCP were tested over a very limited number of projects, thereby reducing the credibility of the models. To avoid that, we used two datasets collected from software organizations and university students [10]. The first dataset (hereafter DS1) contains forty-five projects developed for information systems projects such as chains of hotels, multi-branch universities, and multi-warehouses bookstores. The architecture used to develop these projects are a 2-tier desktop application and 3-tier web architecture [10]. The second dataset (DS2) containing 65 projects were collected from 4[th] year and Master's degree students at a University in North America. The projects were developed and implemented using UML diagrams and object-oriented programming languages [10]. Tables 4 and 5 show the descriptive statistics of the datasets. Both datasets are considered homogenous since the projects in each dataset belong to the same origin. For example, the DS1 was collected from industrial projects. Interestingly, all datasets shared the same feature description, which enabled us to build up other datasets that combined all projects. So we merged DS1 and DS2 into one dataset, called DS3 hereafter. This enables us to investigate the usefulness of environmental factors to predict productivity over the heterogeneous dataset. The statistical description of DS3 is presented in Table 6.

TABLE 4. Descriptive statistics of DS1 dataset

| Variable | Mean | StDev | Min | Median | Max | Skewness | Kurtosis |
|---|---|---|---|---|---|---|---|
| UCP | 739.3 | 1563.9 | 33 | 154 | 7027 | 3.0 | 11.7 |
| Effort | 20573.5 | 47326.9 | 570 | 3248 | 224890 | 3.2 | 12.4 |
| productivity | 24.1 | 5.1 | 14 | 24 | 33 | 0.0 | 2.2 |

TABLE 5. Descriptive statistics of DS2 dataset

| Variable | Mean | StDev | Min | Median | Max | Skewness | Kurtosis |
|---|---|---|---|---|---|---|---|
| UCP | 82.6 | 20.7 | 40.0 | 81.0 | 149.0 | 0.8 | 4.1 |
| Effort | 1672.4 | 414.3 | 696.0 | 1653.0 | 2444.0 | -0.1 | 2.2 |
| productivity | 20.8 | 4.8 | 11.0 | 21.0 | 32.0 | 0.2 | 2.7 |

TABLE 6. Descriptive statistics of DS3 dataset

| Variable | Mean | StDev | Min | Median | Max | Skewness | Kurtosis |
|---|---|---|---|---|---|---|---|
| UCP | 351.3 | 1045.3 | 33.0 | 95.0 | 7027.0 | 5.1 | 30.1 |
| Effort | 9404.7 | 31486.6 | 570.0 | 2140.0 | 224890.0 | 5.3 | 32.1 |
| productivity | 22.1 | 5.2 | 11.0 | 23.0 | 33.0 | 0.1 | 2.5 |

Commenting on the statistical properties of the datasets that were employed, we notice that the productivity in all datasets followed a normal distribution with skewness very close to zero. This has also been confirmed in Figures 4 to 6, respectively. The productivity distribution of DS2 is flatter than normal and has kurtosis values of 2.7, while it is more sharply peaked than normal for DS1. Although this is quite intuitive from the histograms, it was also confirmed by the D'Agostino Pearson test for normality. The Kurtosis is a measure of how outlier-prone a distribution is. The kurtosis of the normal distribution is 3. Distributions that are less outlier-prone than the normal distribution have a kurtosis of less than 3. In all datasets, the kurtosis values were less than 3, which confirms that the distribution of productivity is less outlier-prone. From the comparison between DS1 and DS2, in terms of productivity, we can notice that the productivity of student projects in DS2 is slightly smaller in average than industrial projects, and the distribution of productivity for DS2 is smaller than that of DS1. Thus, it is important from the manager's point of view to recognize that the projects in DS2 need less productivity, perhaps due to the type of projects that were developed in university as graduation projects, which are usually restricted by time and are less complex.

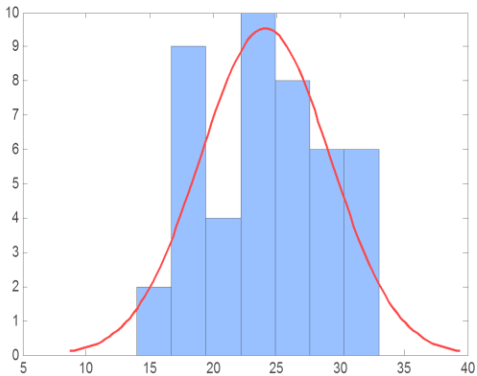 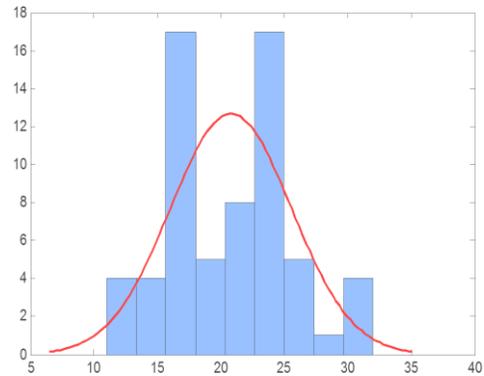

Figure 4. Productivity histogram of DS1        Figure 5. Productivity histogram for DS2

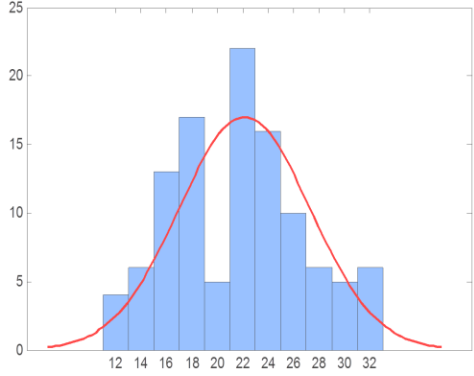

Figure 6. Productivity histogram of DS3

## 7 Results and Discussion

The purpose of the established studies was to investigate the relationship between productivity and environmental factors of the UCP method with the aim to build productivity prediction models. The outcome of these models will be used later to predict the effort from UCP at an early stage. As discussed in Section 2, the eight environmental factors provide a potential indicator for the amount of project productivity needed to accomplish a software project. If we look closer at these factors, we can see that each one provides significant information about project productivity. Nevertheless, it has been reported that the assessment of environmental factors is subject to manager experience when they use a 0-5 ordinal scale, which can have a negative impact on the final estimate. Schneider and Winters [16] and Nassif et al. [10] attempted to exclude environmental factors from calculating and adjusting UCP and use them only for determining productivity ratios. Their approaches are useful when an organization has no historical data and the margin of error is acceptable. So one should be careful about the sensitivity of the ordinal scale of adjustment factors. However, the availability of a large number of projects allowed us to make further investigation into the relationship between productivity and environmental factors. In the first study, we built a classification model based on the basic procedure of the S&W model. The outcome model of this study was called M1. Similarly, in the second study we designed another classification model based on environmental factors and productivity by creating initial productivity labels from the k-means clustering algorithm. The outcome model of study two is called M2. The third study built a stepwise regression model between environmental factors and productivity called M3. Lastly, the fourth study used an analogy-based estimation with R2M to predict and adjust productivity from environmental factors. The outcome model of the fourth study is called M4. The productivity estimated by these models can be used with UCP as predictors to estimate the early effort, as shown in Equation 1. As the first step of validation, we needed to ensure that the generated models were meaningfully predicted. To accomplish that, we used the validation framework proposed by Shepperd and MacDonell [40], which is based on SA and effect size. Table 7 shows the SA and effect size for the four models over three datasets. The models with large SA and effect size are regarded as predicting models; otherwise, the prediction has likely arisen by chance.

TABLE 7. SA and Δ analysis considering random guessing model as baseline

| Dataset | | Karner | S&W | Nassif | M1 | M2 | M3 | M4 |
|---|---|---|---|---|---|---|---|---|
| DS1 | SA% | 81.13 | 87.99 | 86.37 | 93.2 | 93.2 | 92.9 | 91.80 |
| | Δ | 0.44 | 0.48 | 0.47 | 0.51 | 0.51 | 0.51 | 0.5 |
| DS2 | SA% | 31.07 | 6.48 | 31.78 | 36.4 | 37.7 | 42.2 | 32.88 |
| | Δ | 0.44 | 0.09 | 0.45 | 0.52 | 0.54 | 0.6 | 0.47 |
| DS3 | SA% | 81.6 | 87.4 | 86.4 | 88.7 | 88.6 | 87.5 | 86.7 |

|   |   |   |   |   |   |   |   |
|---|---|---|---|---|---|---|---|
|   | Δ | 0.28 | 0.3 | 0.3 | 0.29 | 0.31 | 0.31 | 0.3 |

Interestingly, we can comfortably confirm that all proposed models produce very high SA over datasets DS1 and DS3, which means that they are meaningfully predicting and better than random guessing. The obtained SA over DS2 is not high, but we can judge that this model can work better than guessing. The good enhancements obtained in DS1 and DS3 can be attributed to several reasons. First, the dataset DS1 contains industrial projects that were collected by experts, so the quality of data collection is high. Second, the dataset DS2 only contains student projects that are regarded as simple and of low quality and have enough time to be delivered. On the other hand, the DS3 dataset, which contains both student and industrial projects from DS1 and DS2, still produced a higher SA than DS2. The effect size test of all models demonstrated considerably large effect sizes over DS1 and DS2, which confirmed a great effect improvement against guessing (i.e. Δ ≈ 0.5). Surprisingly, the effect size results on DS3 for all models were not high, perhaps due to the effect of low quality data on high quality data during learning or predicting the procedure. From the initial observation, we can recommend that there should be mainly reliance on the datasets that come from organizations since their projects were usually collected by experts who have sufficient knowledge in objectory.

Table 8. SA and Δ analysis considering Default model as baseline

| Dataset |     | S&W    | Nassif | M1   | M2   | M3   | M4   |
|---------|-----|--------|--------|------|------|------|------|
| DS1     | SA% | 36.38  | 27.79  | 63.3 | 65.3 | 62.4 | 57.3 |
|         | Δ   | 0.14   | 0.10   | 0.24 | 0.25 | 0.23 | 0.22 |
| DS2     | SA% | -35.68 | 1.03   | 7.45 | 9.5  | 16.2 | 3.3  |
|         | Δ   | 0.49   | 0.01   | 0.1  | 0.13 | 0.23 | 0.04 |
| DS3     | SA% | 31.2   | 25.9   | 14.0 | 38.7 | 32.1 | 27.3 |
|         | Δ   | 0.07   | 0.06   | 0.1  | 0.1  | 0.08 | 0.07 |

Similarly, we needed to measure the amount of accuracy improvements that have been achieved by the proposed four models against the Karner model. The Karner model is considered the original baseline model that presumes the productivity ratio is similar for all projects, which is relatively 20 person-hours/UCP. So we re-ran the SA and effect size test using the Karner model as the baseline model. Table 8 presents the SA and effect size for all models over all datasets, using the Karner model as the baseline. Interestingly, all four proposed models produced relatively similar improvements that vary from one dataset to another. For example, they produced great improvements against the Karner model over the DS1 dataset, fair improvements over the DS3 dataset, and few improvements over the DS2 dataset. We believe this can be attributed to the quality of data in DS2. So we believe that using the Karner model for a student project would work well for that project but not necessarily be superior.

TABLE 9. MAE results

| Dataset | Karner  | S&W     | Nassif  | M1     | M2     | M3     | M4     |
|---------|---------|---------|---------|--------|--------|--------|--------|
| DS1     | 6120.62 | 3893.73 | 4419.96 | **2231.8** | 2402.0 | 2300.8 | 2614.3 |
| DS2     | 329.04  | 446.42  | 325.64  | 281.7  | 287.1  | **275.6** | 318.0  |
| DS3     | 2698.3  | 1856.7  | 2000.6  | 1537.1 | **1654.1** | 1832.2 | 1961.3 |

TABLE 10. MBRE results

| Dataset | Karner | S&W   | Nassif | M1   | M2   | M3       | M4   |
|---------|--------|-------|--------|------|------|----------|------|
| DS1     | 28.22  | 21.46 | 25.00  | 15.9 | 16.6 | **15.5** | 19.4 |
| DS2     | 23.18  | 30.36 | 24.28  | 19.9 | 20.2 | **19.3** | 22.4 |
| DS3     | 25.2   | 26.7  | 24.6   | 20.3 | 20.2 | **19.6** | 22.7 |

TABLE 11. MIBRE results

| Dataset | Karner | S&W | Nassif | M1 | M2 | M3 | M4 |
|---|---|---|---|---|---|---|---|
| DS1 | 20.72 | 16.67 | 18.47 | 12.5 | 12.8 | **12.1** | 14.9 |
| DS2 | 17.40 | 19.70 | 17.78 | 14.7 | 14.9 | **14.6** | 16.8 |
| DS3 | 18.8 | 18.5 | 18.1 | 14.9 | 14.9 | **14.4** | 16.4 |

Tables 9-11 present the predictive accuracy in terms of MAE, MBRE, and MIBRE. In each table we compared the proposed four models against the previous prediction models that used productivity with UCP as predictor variables for effort estimation. The cells with grey highlight and boldface font represent the best accuracy across each dataset. Interestingly, all proposed models work comparatively well over all datasets, and better than previous models. Specifically, we noticed that the M3 that uses stepwise regression is slightly better than other models (M1, M2, and M4). Nevertheless, we also noticed that both M1 and M2, using classification algorithm, work better than R2M. This is an indication of the superiority of the learning capability of the classification algorithm on environmental factors. The Karner and S&W models produced the worst performance in comparison to Nassif and R2M. The S&W model produced the worst estimates with a large error deviation, as confirmed in MAE results. Another important finding was that relying on the structure of data to classify project productivity, as in M1 and M2, is significantly better than using expert classification, as in the S&W or Nassif model. All these improvements can be partially attributed to the strong correlation between effort and UCP, as shown in Figure 7. It is obvious from the figure that the relationship between effort and UCP is quite linear with correlation a coefficient equal to 0.838 and a p-value less than 0.01. Above all, we can comfortably confirm the importance of environmental factors in predicting productivity and, consequently, effort.

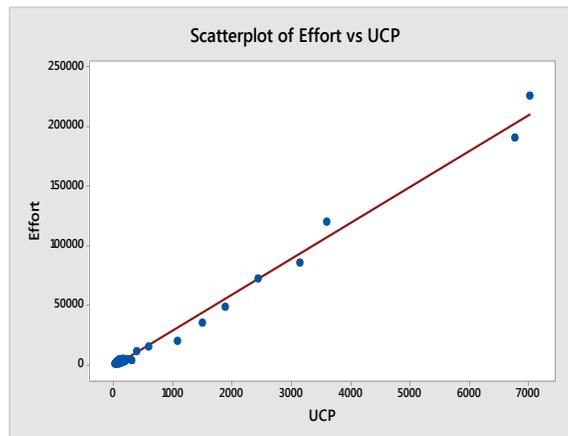

Figure 7. Relationship between Effort and UCP over DS3

The results without statistical significance are not justified, therefore, we ran the Wilcoxon sum rank statistical significance test based on model residuals, at the significance level of 0.05. We chose the Wilcoxon test as a non-parametric statistically significant test because the data used were not normally distributed and, thus, parametric tests such as ANOVA and t-test cannot be used. The results of our comparison are shown in Tables 12 to 14. Table 12 shows that models are statistically significant against the Karner model, which indicates that they produce better accuracy than the default model. Also, we can notice that there is no significant difference between all proposed models (i.e., M1 to M4). On the other hand, the proposed four models yielded significantly different predictions in comparison to the S&W and Nassif models over DS2 dataset, but not more significant than Karner. This suggests that the projects in which the estimation was made by students tended to be very close to the Karner model. Similar results were obtained over DS3 but they were superior to M3, as they yielded different predictions than all other models.

TABLE 12. Statistical significance of test results over DS1

|  | Karner | S&W | Nassif | M1 | M2 | M3 |
|---|---|---|---|---|---|---|
| S&W | 0.001 | | | | | |
| Nassif | 0.039 | 0.188 | | | | |
| M1 | 0.003 | 0.04 | 0.02 | | | |
| M2 | 0.004 | 0.03 | 0.04 | 0.936 | | |
| M3 | 0.007 | 0.04 | 0.03 | 0.317 | 0.388 | |
| M4 | 0.001 | 0.04 | 0.03 | 0.929 | 0.949 | 0.362 |

TABLE 13. Statistical significance of test results over DS2

|  | Karner | S&W | Nassif | M1 | M2 | M3 |
|---|---|---|---|---|---|---|
| S&W | 0.000 | | | | | |
| Nassif | 0.013 | 0.000 | | | | |
| M1 | 0.375 | 0.005 | 0.004 | | | |
| M2 | 0.419 | 0.004 | 0.002 | 0.967 | | |
| M3 | 0.414 | 0.000 | 0.000 | 0.911 | 0.885 | |
| M4 | 0.981 | 0.000 | 0.008 | 0.397 | 0.402 | 0.478 |

TABLE 14. Statistical significance of test results over DS3

|  | Karner | S&W | Nassif | M1 | M2 | M3 |
|---|---|---|---|---|---|---|
| S&W | 0.000 | | | | | |
| Nassif | 0.743 | 0.000 | | | | |
| M1 | 0.000 | 0.07 | 0.001 | | | |
| M2 | 0.000 | 0.09 | 0.001 | 0.952 | | |
| M3 | 0.038 | 0.000 | 0.006 | 0.012 | 0.009 | |
| M4 | 0.001 | 0.048 | 0.000 | 0.169 | 0.127 | 0.129 |

Other results offered yet more evidence of the performance of the proposed prediction models over previous models. Figures 8 to 10 show the Scott-Knott test analysis [42] among all models over each single dataset. The Scott-Knott is a multiple comparison statistical method based on the idea of clustering where the criteria of clustering is the significance test between the absolute errors of the methods. To use the Scott-Knott test, the absolute errors must be normally distributed, otherwise they must be transformed to a normal distribution using a transformation algorithm, such as the Box-Cox method [42]. The Scott-Knott test uses one-way analysis of variance (one way ANOVA) which tests the null hypothesis that the methods under comparison are not statistically different against the alternative hypothesis and that the methods can be partitioned into subgroups. The principal reason for using the Scott-Knott method is attributed to its ability to separate the prediction models into non-overlapping groups. We tested all absolute residuals in all datasets using the D'Agostino Pearson test, and we found that they were not normally distributed. Therefore, all absolute errors were transformed using the Box-Cox method. In Figures 8 to 10, the x-axis represents prediction model names sorted according to their ranks where better places start from the right hand side. The y-axis represents the transformed absolute errors and the small circles on each vertical line represent the mean of transformed absolute errors. From the figures we notice that the proposed four models (M1 to M4) are grouped together in one cluster and ranked higher than other models over the DS1 dataset. The main observation from these figures is that there is no consistency among them regarding which prediction model is superior. For example, we can see that M1 ranks first over DS1, as shown in Figure 5, where it ranks 2 over DS2 and DS3. This emphasizes the great effect of data quality on the final outcome.

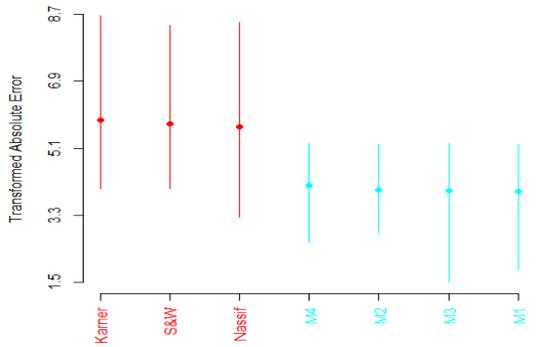

Figure 8. Plot of the Scott-Knott algorithm based on Transformed absolute errors over DS1

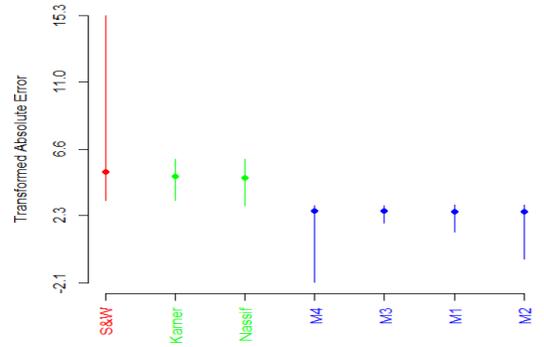

Figure 9. Plot of the Scott-Knott algorithm based on Transformed absolute errors over DS2

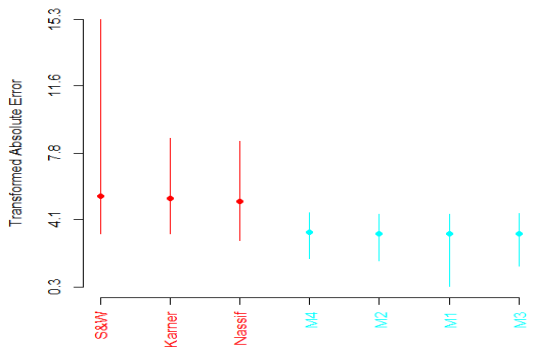

Figure 10. Plot of the Scott-Knott algorithm based on Transformed absolute errors over DS3

The success of any prediction model can be viewed as three dimensional points: accuracy, stability, and transparency [37]. Accuracy is an important factor for judging the superiority of any prediction model, in terms of software effort estimation, as managers need to rely on accurate decisions not wrong ones [37]. The stability of a prediction model across different conditions is also important because a practitioner would be more confident of the outcome if he/she feels that the model used can work comfortably under different experimental conditions [37]. Finally, transparency (i.e., how easy it is to understand the model generated by the approach) is also required, since no one can trust the final results without knowing what goes inside the model [37]. However, our study shows that all proposed four prediction models (M1 to M4) are in general more stable than other models that occupy the first places over each dataset, as confirmed by a Scott-Knott test analysis. This facilitates and allows the manager to choose any model among M1, M2, M3, and M4, as they all produce meaningful and accurate predictions. Nevertheless, some models, such as M1 and M2, may be less transparent than others, so it would be up to the practitioner to decide what to use. The decision rests on whether the manager prefers a model that has been shown to achieve better accuracy and stability but is less transparent over all or a model that has shown to be less accurate and less stable, but is more transparent. Perhaps, using the Ensemble method could be an efficient solution for such a decision. To further investigate the stability of the proposed four models we conducted a win-tie-loss method [43], which compares all prediction models used across all datasets and uses different accuracy measures, as shown in Figure 11. For each pair of the model, the algorithm first checks if two models $Model_i$; $Model_j$ are statistically different according to the Wilcoxon sum rank test; otherwise, we increase $tie_i$ and $tie_j$. If the error distributions are statistically different, we update $win_i$; $win_j$ and $loss_i$; $loss_j$, after checking which one is best, according to the performance measure at hand $E$. The performance measures used here are *MIBRE*, *MBRE*, and *MAE*.

```
1    win_i ← 0, tie_i ← 0, loss_i ← 0
2    win_j ← 0, tie_j ← 0; loss_j ← 0
3    if WILCOXON(MAE (M_i), MAE(M_j), 95) says they are the same then
4        tie_i ← tie_i + 1;
5        tie_j ← tie_j + 1;
6    else
7        if better(E(M_i), E(M_j)) then
8            win_i ← win_i + 1
9            loss_j ← loss_j + 1
10       else
11           win_j ← win_j + 1
12           loss_i ← loss_i + 1
13       end if
14   end if
```

Figure 11. Pseudo code for *win, tie, loss* calculation between $Model_i$ and $Model_j$ based on performance measure $E$ [43].

Figures 12 to 15 show the results of the sum of *win*, *tie*, and *loss* for all models across all datasets used. The *tie* result is not informative enough to distinguish among models, therefore, we depended on the results of *win* and *loss*, which would be more informative. If we look closer at the first rank in each dataset, it is impossible to identify any common methods among them, so we cannot claim that there is a winner across all datasets. An important observation is the ranking instability across all datasets. It is clear that there was no stable ranking across all datasets. However, although there is ranking instability, it is crucial to obtain the important information that is hidden in this lack of convergence and to interpret the findings with caution. Remarkably, all proposed models obtain zero losses, which indicate they work comparatively well and surpass the previous models. Among them, M1, M2, and M3 have achieved a large number of wins. The Karner model achieved the worst results.

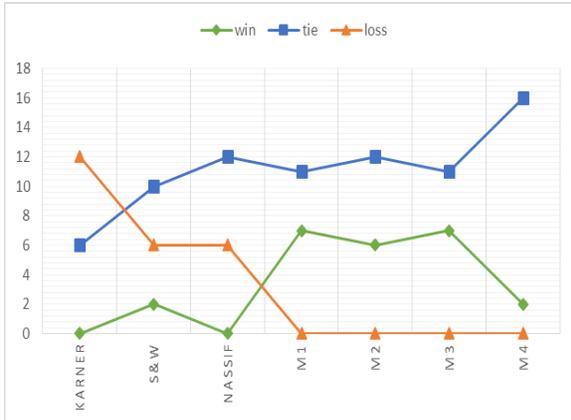

Figure 12. Win-tie-loss results over DS1

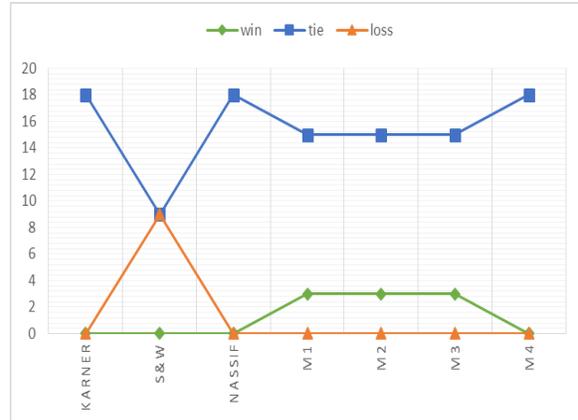

Figure 13. Win-tie-loss results over DS2

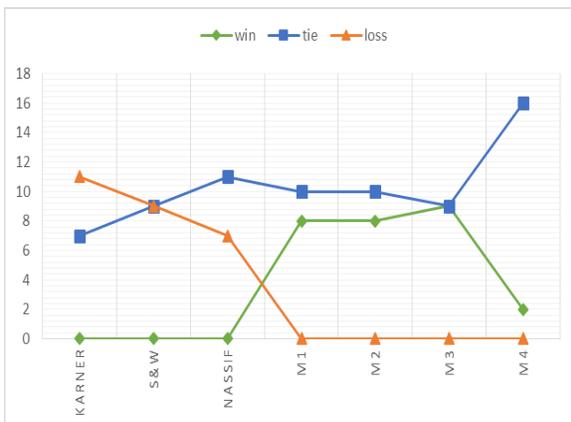

Figure 14. Win-tie-loss results over DS3

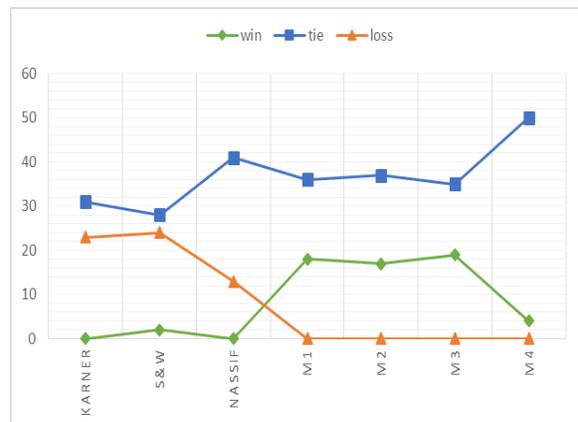

Figure 15. Accumulative win-tie-loss results over all datasets

Commenting on the above results we can see that learning productivity from environmental factors is more efficient that using expert assumptions. Additionally, using dynamic productivity values, as in the four models, produces more accurate and significant results than using static or very limited values as in the S&W and Karner models. Finally, the environmental factors for each individual project must be assessed when all information about the project is readily available because bad assessment may enforce the proposed models to behave differently from organization to organization.

## 8    Threats to Validity

In this section we mention some threats that might have affected the validity of the proposed models. We divide the threats into internal, external, and construct.

1) *Internal validity* is the degree to which conclusions can be drawn with regard to the configuration setup. First, there is no outstanding approach to discover the optimum number of clusters before running the k-means clustering algorithm. So we used a clustering validity measure to help us in determining the number of clusters. This measure is based on minimizing the distance between observations within clusters and increasing the distance among cluster centers. Although there are plenty of validity measures, we believe that the measure used is enough to give us an indication of how many clusters each training dataset needs. Regarding the validation procedure, we used leave-one-out cross validation because it produces lower bias estimates and higher variance than n-Fold cross validation.

2) *Construct validity* ensures that we are measuring what we actually intended to measure. Despite the widespread use of MRE and its derived measures in comparison among prediction models, we did not use them in this study because they have been criticized as being biased. However, we used more robust evaluation measures such SA, effect size, MBRE, and MIBRE, which do not produce an asymmetric error distribution.
3) *External validity* is the ability to generalize the findings obtained from our comparative studies. In this study, we used two datasets that are not public. This might affect the repeatability of the proposed models in the future.

## 9 Conclusions and Future Work

Project productivity is a key factor for producing effort estimates from UCP when a historical dataset is absent. This paper examined two important research questions related to project productivity, as mentioned in Section 1. To answer RQ1, four studies were proposed to examine this issue. The proposed models by those studies could work in the presence and absence of historical data because they have been constructed from a large number of observations and use environmental factors as input. These factors can be easily determined at the early stages, but one should be careful about the assessment of these factors because they are subject to the experience of people who use 0-5 ordinal scale. This can have a negative impact on the consistency of estimates.

The results obtained by these studies showed that there is a good relationship between environmental factors and project productivity based on the estimation accuracy obtained. Specifically, we can see that the M3 model that uses stepwise regression was the most accurate with significant results. Regarding the datasets, we observed that all proposed models work better over a DS1 dataset that comes from industry. The quality of data collection also has a significant impact on the predictive accuracy, as shown in the results. We also noticed that all proposed models work better than previous models over homogenous and heterogeneous datasets.

The enhancement achieved by M1 and M2 results can be attributed to a number of reasons. Most UCP datasets have a high granularity of labeling the data. So the outcome of the class decompositions that occurred in M1 and M2 was a fine-grained labeled dataset. As we can see in S&W and Nassif, only three or four class labels were used to classify dataset. Even with suitable class labeling, inherent subclasses can be often discovered at a later stage. This is true in the software domain. For example, managers can start with two class labels (high productivity and low productivity), in which each class would be decomposed to more coherent class labels, such as fairly low, very low, extremely low, and the same decomposition used for a high productivity label. Consequentially, more precision can be enhanced if a class decomposition of the productivity is applied, resulting in a multi-class problem. Finally, early effort estimation is a complex process with a high degree of uncertainty and non-linearity. Class-decomposed data can simplify the process by looking at cohesive subclasses, instead of modeling a more complex higher granular set of classes.

Regarding RQ2: We have noticed that the proposed models M1 to M4, in the four studies, yield dynamic productivity ratios. These dynamic productivity values contributed to better accuracy in terms of MAE, MBRE, and MIBRE more than static values as in Karner and S&W models. To conclude, we can confirm that the productivity can be predicted efficiently from environmental factors. We recommend excluding environmental factors from computing UCP and making them available for predicting productivity. The productivity predicted can also work well with UCP to estimate early effort, so there is no need to build complex effort prediction models. To further improve the models, we encourage project manager to better understand how to assess and evaluate the environmental factors, as they have significant impact on productivity prediction.

Future work will focus on collecting and using more datasets in order to generalize our findings. Moreover, we will conduct more experiments to compare between the proposed approach and models when all UCP variables are taken into consideration. Furthermore, we will experiment the effect of TCF, in addition to ECF, on productivity.

## Acknowledgement

Dr. Ali Bou Nassif would like to thank the University of Sharjah for supporting this research.